\pdfoutput=1
\documentclass[journal, 10pt]{IEEEtran}
\IEEEoverridecommandlockouts
\usepackage[utf8]{inputenc}
\usepackage{cite}
\usepackage{amsmath,amsfonts, amsthm}
\usepackage{algorithmic}
\usepackage{graphicx}
\usepackage{textcomp}
\usepackage{xcolor}
\usepackage[hidelinks]{hyperref}
\usepackage{subcaption}
\usepackage{pdfpages}
\usepackage{siunitx}
\usepackage[normalem]{ulem}
\theoremstyle{definition}

\usepackage{svg}

\def\BibTeX{{\rm B\kern-.05em{\sc i\kern-.025em b}\kern-.08em
 T\kern-.1667em\lower.7ex\hbox{E}\kern-.125emX}}

\begin{document}
\title{Developing an Underwater Network of Ocean Observation Systems with Digital Twin Prototypes - A Field Report from the Baltic Sea}

\author{Alexander~Barbie, Niklas~Pech, Wilhelm~Hasselbring, Sascha~Fl\"ogel, Frank~Wenzh\"ofer, Michael~Walter,~\IEEEmembership{Senior Member,~IEEE}, Elena~Shchekinova, Marc~Busse, Matthias~Türk, Michael~Hofbauer, Stefan~Sommer

\thanks{A. Barbie, F. Wenzhöfer, M. Hofbauer are with the Alfred Wegener Institute Helmholtz Centre for Polar and Marine Research, 27570 Bremerhaven, Germany (e-mail: \{alexander.barbie, frank.wenzhoefer, michael.hofbauer\}@awi.de)}
\thanks{A. Barbie, N. Pech, S. Flögel, E. Shchekinova, M. Busse, M. Türk are with the GEOMAR Helmholtz Centre for Ocean Research Kiel, 24148 Kiel, Germany (e-mail: \{abarbie, npech, sfloegel, esheckinova, mbusse,mtuerk, ssommer\}@geomar.de)}
\thanks{A. Barbie, W. Hasselbring are with the Christian-Albrecht University, Software Engineering Group, 24118 Kiel, Germany (e-mail: hasselbring@email.uni-kiel.de)}
\thanks{F. Wenzhöfer, M. Hofbauer are with the Max Planck Institute for Marine Microbiology, HGF-MPG Joint Research Group for Deep-Sea Ecology and Technology, 28359 Bremen, Germany}
\thanks{M. Walter is with the German Aerospace Center (DLR), Institute of Communications and Navigation, 82234 Wessling, Germany (e-mail: m.walter@dlr.de)}}

\maketitle

\begin{abstract}
During the research cruise AL547 with RV ALKOR (October 20-31, 2020), a collaborative underwater network of ocean observation systems was deployed in Boknis Eck (SW Baltic Sea, German exclusive economic zone (EEZ)) in the context of the project ARCHES (Autonomous Robotic Networks to Help Modern Societies). This network was realized via a Digital Twin Prototype approach. During that period different scenarios were executed to demonstrate the feasibility of Digital Twins in an extreme environment such as underwater. One of the scenarios showed the collaboration of stage IV Digital Twins with their physical counterparts on the seafloor. This way, we address the research question, whether Digital Twins represent a feasible approach to operate mobile ad hoc networks for ocean and coastal observation.
\end{abstract}

\begin{IEEEkeywords}
Baltic Sea, Digital Twins, Field Report, Ocean Observation, Prototyping, Robotic Networks
\end{IEEEkeywords}

\section{INTRODUCTION}\label{sec:introduction}
The ocean is the largest ecosystem on earth facing dramatic changes such as deoxygenation, warming, acidification, and contamination by industrial pollution, to name a few. The involved physical, biological and biogeochemical processes are highly diverse and form a complex network of causal interactions that act on variable temporal and spatial scales. Some of them are fast (less than a day) and locally confined (several \SI{100}{m} to km) such as the response of coastal ecosystems to the passage of storms, input of increased nutrient loads from the coastal drainage area subsequent to heavy rainfall, or locally enhanced primary production effecting the local carbon cycling.
To understand major drivers and ecosystem response in space and time conventional ship based observation programs are not sufficient as ship based operations involve long planning periods and follow a fixed time schedule, which often do not allow to capture sporadic environmental events that are difficult to forecast. 

Despite the availability of suitable carrier platforms (e.g. landers, floats, gliders, waveglider) and the increasing numbers of physical and biogeochemical sensors for marine observation, still many areas in the ocean and along the coasts are undersampled and processes are not quantified adequately.
To simultaneously monitor changes in the water column and at the seafloor an underwater (UW) – robotic sensing network has been developed within the framework of the Helmholtz innovation project ARCHES involving a consortium of partners from AWI (Alfred-Wegener-Institute Helmholtz Centre for Polar and Marine Research, Bremerhaven, Germany), DLR (German Aerospace Center, Oberpfaffenhofen, Germany) and the GEOMAR (Helmholtz Centre for Ocean Research Kiel, Germany). The major aim was to implement a robotic sensing network, which is able to autonomously respond to changes in the environment by adopting its measurement strategy. 
During the research cruise AL547 with RV ALKOR (20.10. – 31.10.2020) the functionality of the network has been demonstrated. It was aimed to establish a mobile ad hoc network (MANET) \cite{manets_survey_2018} of heterogeneous, autonomous and interconnected robotic systems. The network consists of five different stationary and mobile sensing platforms that exchange environmental data with each other and the research vessel.
Despite the technological focus as a science case, the research cruise addressed the oxygen dynamics in coastal ecosystems, which are increasingly impacted by eutrophication and the progressing loss of oxygen (hypoxia) \cite{breitburg2018declining}. Particularly the deeper part of Eckernförder Bay experiences seasonal hypoxia, yet strong fluctuations of oxygen in the bottom water also occur at various frequencies (for reference see web page of the Boknis Eck time series station: https://www.bokniseck.de/de/literature). Hence, beside other sensors implemented in the different sensing platforms all of them were equipped with at least one oxygen sensor and only data of the oxygen sensors were transferred among the various network participants. 

Here, we report on the experience of employing Digital Twins in the above described field experiment. \autoref{sec:context}, briefly describes the set-up of the network, the Digital Twin Prototype approach is introduced in \autoref{sec-dtp}. In \autoref{sec-field} we report on the field experiment. Conclusions and future work are discussed in \autoref{sec:conclusion}.

\begin{figure*}[tb]
\centering
\begin{subfigure}{0.3\textwidth}
 \centering
 \includegraphics[width=\textwidth]{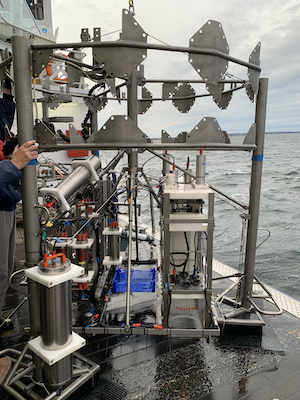}
 \caption{BIGO Lander}
 \label{fig:bigo}
\end{subfigure}
\hspace{0.1cm}
\begin{subfigure}{0.3\textwidth}
 \centering
 \includegraphics[width=\textwidth]{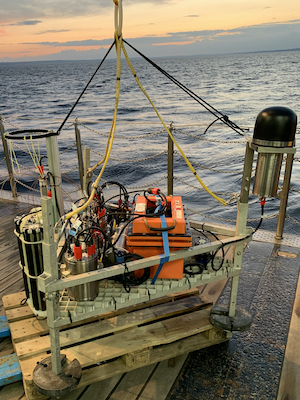}
 \caption{CRAWLERSIM}
 \label{fig:crawlersim}
\end{subfigure}
\hspace{0.1cm}
\begin{subfigure}{0.3\textwidth}
 \centering
 \includegraphics[width=\textwidth]{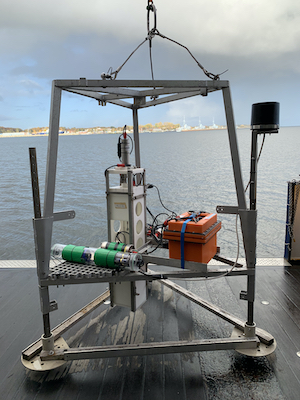}
 \caption{FLUX Lander}
 \label{fig:flux}
\end{subfigure}
\\
\vspace{0.5cm}
\centering
\begin{subfigure}{0.47\textwidth}
 \centering
 \includegraphics[width=1.0\textwidth]{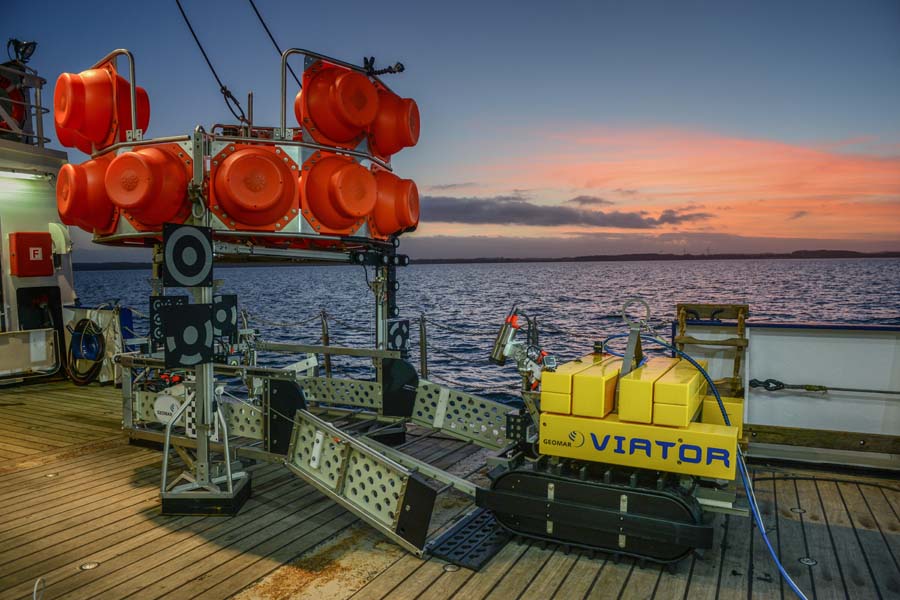}
 \caption{MANSIO \& VIATOR}
 \label{fig:mansioviator}
\end{subfigure}
\caption{The ocean observation systems participating in the ARCHES network.}
\label{fig:multiplelander}
\end{figure*}

\section{PROJECT CONTEXT}\label{sec:context}
Since each field test of ocean observation systems always requires a research vessel and thus, a crew, it is quite expensive to test networks of observation systems this way. Furthermore, the placement or retrieval of an observation system from the seafloor, which would happen every time a software error has to be fixed, takes up to several hours, depending on the deployment depth and weather conditions. The advancement in hardware and software technology and especially in the embedded software domain, enables the usage of “Digital Twins” (DT) to reduce all these time-consuming activities and costs around a field test. In a previous study, the concept of how the Digital Twins in this network had been developed was presented in \cite{MFI2020}, see also \autoref{sec-dtp}.

The actual set up of the network and the geographical setting will be published elsewhere. Briefly, the network was established at two sites in the Eckernförder Bay (western Baltic Sea) close to Kiel. Major focus was on the region at the entrance of the Eckernförde Bay close to the Boknis Eck time series Station located approximately \SI{1}{km} from the coast at the position \ang{54;31.77;} N, \ang{010;02.36;} E in water depths of $17$ to $24$ meters.

The sensing carriers of our network included two lander type platforms (BIGO Lander, FLUX Lander), see Fig.~\ref{fig:multiplelander}, which are used outside of the ARCHES project to measure the exchange of solutes between the seafloor and the water column as well as to monitor changes in the bottom water \cite{sommer2009seabed,bigo2003, wenzhofer2016robex}, see Fig.~\ref{fig:mansioviator}. The CRAWLER SIM system was used to implement the crawler-control software and scientific payload of long-term benthic crawler systems \cite{lemburg2018benthic}. The MANSIO-VIATOR system comprises a stationary lander system serving as hangar (MANSIO) and a mobile deep-sea crawler (VIATOR), Fig.~\ref{fig:mansioviator}. The MANSIO lander serves as a hangar, which is used for transport to the site of investigation and for recovery at the ocean surface as well as to recharge the lithium polymer (LiPo) accumulators on the crawler. 

The transfer of data and commands as well as positioning from all network partners was performed hydroacoustically using Evologics S2C R 7/17 USBL acoustic modems \cite{evologicsmodem717} on each observation system. A shipboard modem was mounted in the moon pool of RV ALKOR. Still the establishment of a reliable and robust hydroacoustic communication is a challenge as summarized by Akyildiz et al. \cite{underwaterchallenge}. In contrast to data transfer using optical systems or electromagnetic waves hydoacoustic communication is slow, it has a high energy demand, small bandwidths, and time-varying multi-path propagation. Furthermore, reliable transmission depends on physical properties of the water column including presence and position of a pycnocline, turbidity, temperature affecting sound velocity. 

\section{DIGITAL TWIN PROTOTYPES}\label{sec-dtp}
In Barbie et al. \cite{MFI2020}, a detailed definition of the ``Digital Twin Prototype'' approach was given based on the Digital Twin definition by Saracco \cite{dtdefinition}. A Digital Twin Prototype is the software/model prototype of a real entity, the Physical Twin. It uses existing recordings of sensing and actuation data over time as digital shadow, to simulate the Physical Twin. In ARCHES each ocean observation system has a corresponding Digital Twin Prototype and Digital Twin. The software on the Physical Twin is identical to the software that is used for the Digital Twin Prototype. The difference is that the Physical Twin has real hardware connected and the Digital Twin Prototype uses emulated hardware components, yet both types synchronize their sensing/actuation data with the corresponding Digital Twin. The Digital Twin differs from the Digital Twin Prototype only by an environment flag that defines it is a Physical Twin or Digital Twin. If the flag is set \emph{true}, the Digital Twin Prototype is a Digital Twin and does not synchronize its sensing/actuation data, instead it receives sensing/actuation data and synchronizes only the incoming commands to its physical counterpart. This approach enables us to start a Digital Twin Prototype and a Digital Twin at the same time in the same development environment and hence, to develop new modules without the need of a connection to the real hardware \cite{MFI2020}.

\begin{figure*}[tb]
 \centering
 \includegraphics[width=1.0\textwidth]{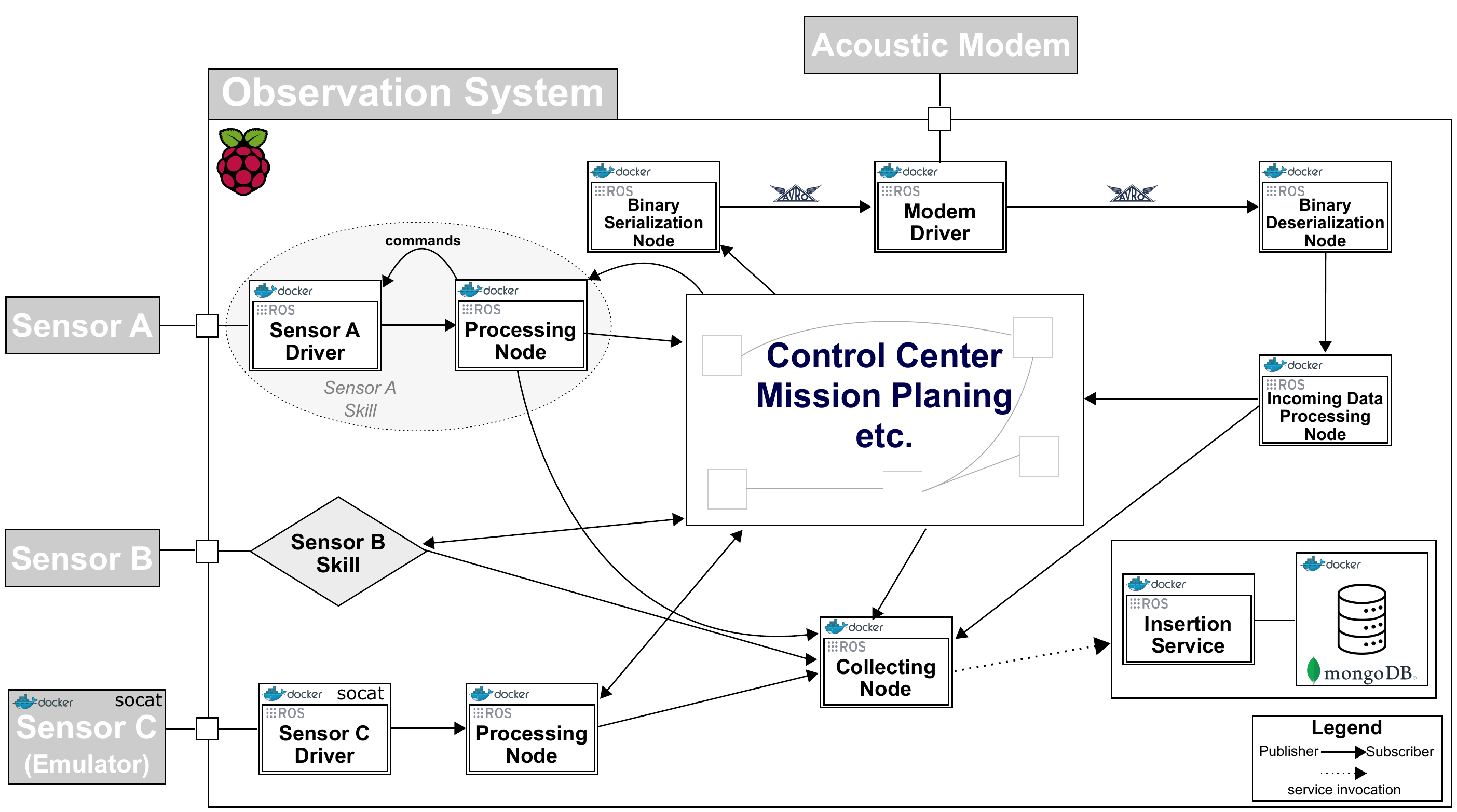}
 \caption{The Software Architecture of the Digital Twins \cite{MFI2020}.}
 \label{fig:architecture}
\end{figure*}

The software framework was developed in Python using the middleware Robot Operating System (ROS) \cite{rospaper}. Each microservice is encapsulated in a Docker container. In Fig.~\ref{fig:architecture}, an overview of the software architecture of the Digital Twins is shown. The Physical Twin side carries real hardware such as sensors/actuators, while on the Digital Twin side the same hardware components are emulated. Even a mix of real and emulated hardware at the same time is possible.

Due to our publish-subscribe architecture, we face some challenges in the synchronization of the Physical Twins and Digital Twins. Messages published by a node XYZ in topic A/B/C on one Twin, have to be synchronized to the corresponding Twin in the same node and the same topic. Nodes that contain publishers or subscribers that are able to synchronize their messages with the Physical or Digital Twin are referred to as \emph{Skill}. Services and drivers can be part of a Skill. Which topics are synchronized with a twin are defined in a list of topics saved on the ROS Parameter Server. This list can contain absolute, relative, and topics mixed with wildcards. There are two types of synchronization. The first, on the Physical Twin, is the transmission of sensor/actuator data and statuses to the Digital Twin. The second, on the Digital Twin, is the transmission of commands to the Physical Twin. The Physical Twin does receive commands only via the corresponding Digital Twin. If a command causes a software failure on the Digital Twin, one can be quite sure that it causes an error on the Physical Twin, too.

The most advanced pair of Physical and Digital Twin in the project ARCHES is the MANSIO-VIATOR system. There are CAD models of both systems in a Gazebo \cite{gazebopaper} simulation. The ROS integration in the Gazebo simulation enables us to connect the models of MANSIO and VIATOR with the ROS software we developed, which is a fundamental idea of Digital Twins. Besides a 3D dynamic multi-robot environment, Gazebo also provides modules that help to simulate more complex hardware such as cameras and lasers. However, this simulation cannot only be used to simulate and visualize the current behavior, which is executed on the Physical Twin, in particular it can be used to develop and test advanced algorithms, e.g., for obstacle avoidance, docking (MANSIO-VIATOR specific), or train AI systems. In combination with our Digital Twin Prototype approach, Gazebo is also part of the automated tests in the continuous delivery pipeline used in ARCHES \cite{MFI2020}.

\section{THE DIGITAL THREAD}
The acoustic modems by Evologics offer different modes for data transmission. Instead of the \emph{Burst Mode} (BM), which has the highest bandwidth with up to \SI{6.9}{kbit/s}, \emph{Instant Messages} (IM) are used. In previous tests they proved to be the most reliable mode for data transmission. Additionally, instant messages can be broadcast to all network participants, which is not possible with the burst mode. The trade off is that the bandwidth for application data is limited to \SI{64}{B/s} with instant messages.

\begin{figure}[tb]
 \centering
 \includegraphics[clip, trim=8cm 0cm 8cm 0cm, width=0.5\textwidth]{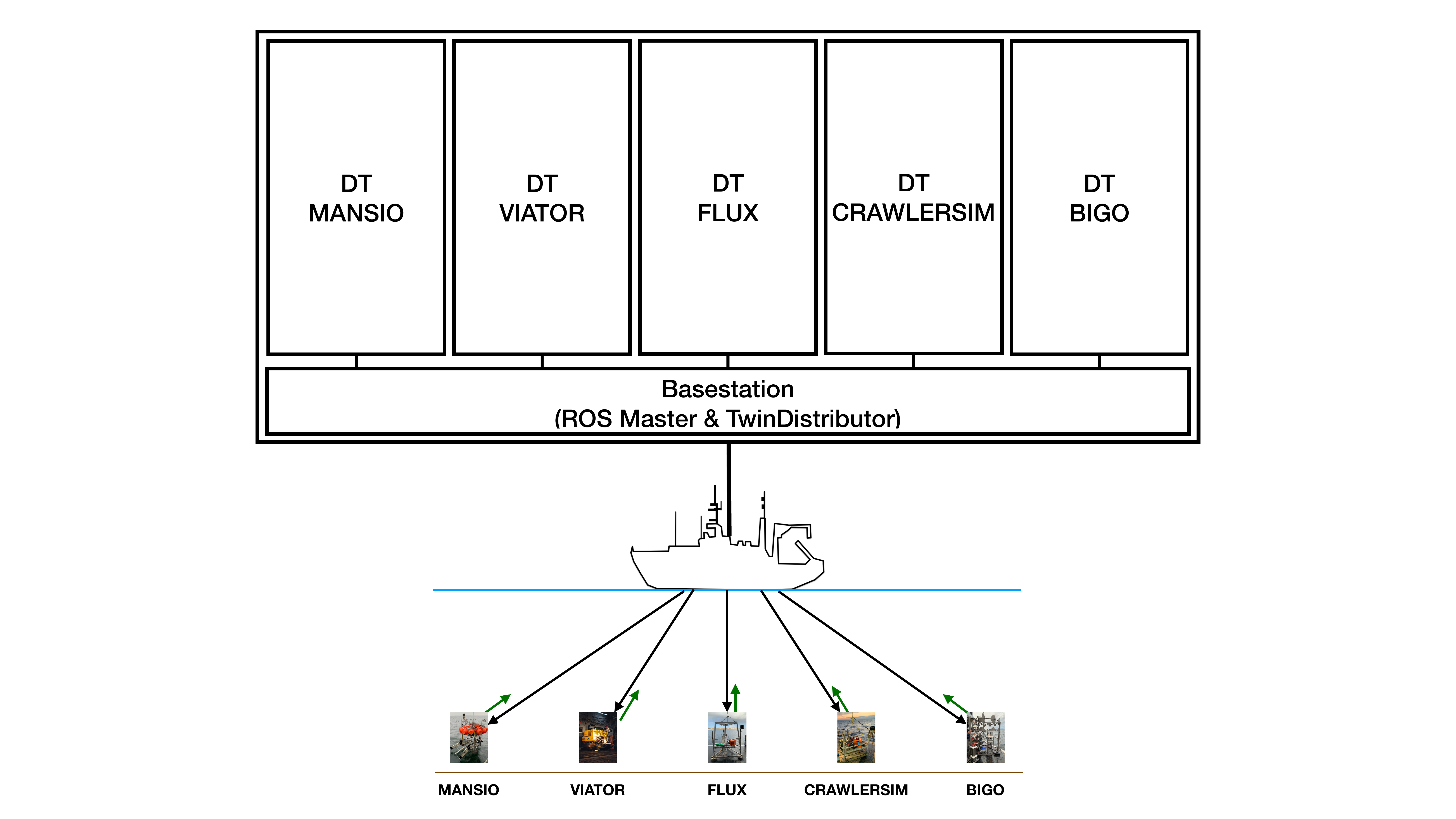}
 \caption{The Network Setup. Refer to Fig.~\ref{fig:multiplelander} for larger pictures of the ocean observation system components.}
 \label{fig:setup}
\end{figure}

Fig.~\ref{fig:setup} shows that each Digital Twin is connected to the \emph{Basestation}, which consists of the ROS Master and one node that distributes the messages of the Physical Twins to the corresponding Digital Twins and vice versa. Connecting the Physical Twin with its digital counterpart on the research vessel, requires a vertical connection from the seafloor to the sea surface and vice versa. Vertical networks, where observation systems on the seafloor send signals to a buoy or research vessel on the sea surface, e.g. the tsunami warning systems GITEWS and DART, were already introduced two decades ago \cite{GITEWS, dart}. However, these networks did only transmit data in a predefined control language that limits the kind of data that can be sent to other platforms. To allow for a more dynamic message exchange, Schneider et al. \cite{dccl} developed a dynamic compact control language (DCCL). The DCCL also addresses another challenge in the underwater domain: the limited bandwidth and thus, the limitation of data that can be transmitted. Serializations in standard data formats such as YAML, which is the format of ROS messages, cause significant overhead. Schneider et al. utilize the binary serialization of Google Protocol Buffers (Protobuf) to serialize object messages and reduce unnecessary overhead. However, in combination with ROS the usage of Protobuf has one disadvantage, Protobuf also uses pre-compiled messages. Hence, two different message definitions by mistake or a missing Protobuf message on one side, leads to a faulty serialization and as a consequence, to corrupted data. To avoid such errors, a new control language was developed in ARCHES utilizing the open-source tool Apache Avro. Avro creates binary de-/serializations without requiring a pre-compile step for code generation, since message schemes can be defined inline. Thus, ROS messages are serialized on the sending observation system to an Avro message and are deserialized on the receiving observation system to a ROS message at runtime. 

To synchronize the Physical Twins and Digital Twins, all ROS messages contain a header that indicates from which Skill and topic a message was published/subscribed. We used ROSBag as storage mechanism during the research cruise. With RQt there is already a Qt-based framework, to inspect all topics in that ROSBag files and visualize the contents in a GUI. In later missions, the main storage will be a MongoDB.

\section{Field Experiment}\label{sec-field}
The goal of this ARCHES Demonstration Mission was to establish a network of heterogeneous and interconnected ocean observation systems and to test its software components. Digital Twins are already used for robotic system in other extreme environments, such as space \cite{nasadt2012}. Hence, one of the research question for this Demo Mission was, whether the Digital Twin approach represents a suitable approach for MANETs in under water applications.

With the Digital Twin Prototype approach we were able to evaluate the scenarios described below in \autoref{subsec:scenarios} in a virtual environment before applying them during the field tests. The Physical Twins were simulated by the Digital Twin Prototypes. Instead of the real acoustic modems, we used the S2C D-MAC Emulator also provided by Evologics. This Demonstration Mission is the first time we evaluated the following scenarios under real conditions. All the tests of the scenarios were performed during the research cruise AL547. The Digital Twins were running on a server on the research vessel ALKOR and shared a common ROS Master provided by the Basestation. The Basestation was connected to the Evologics modem that was located in the moon pool of the research vessel.

\subsection{METHODOLOGY AND SCENARIOS}\label{subsec:scenarios}
\begin{figure*}[tb]
\centering
\begin{subfigure}{0.45\textwidth}
\centering
 \includegraphics[width=\textwidth]{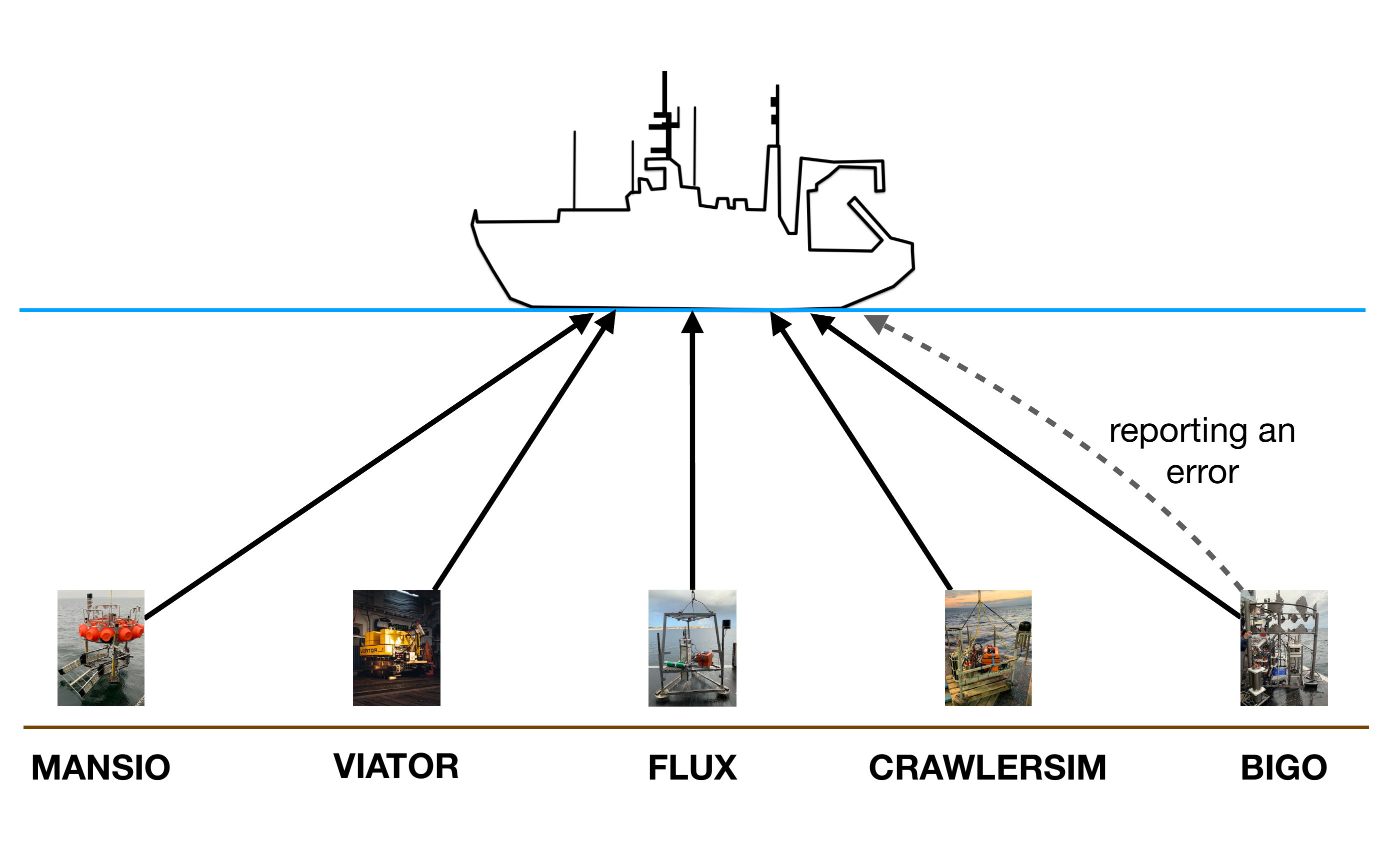}
 \caption{Synchronize the environmental data with the Digital Twins.}
 \label{fig:scenarioa}
\end{subfigure}
\hspace{1cm}
\begin{subfigure}{0.45\textwidth}
\centering
 \includegraphics[width=\textwidth]{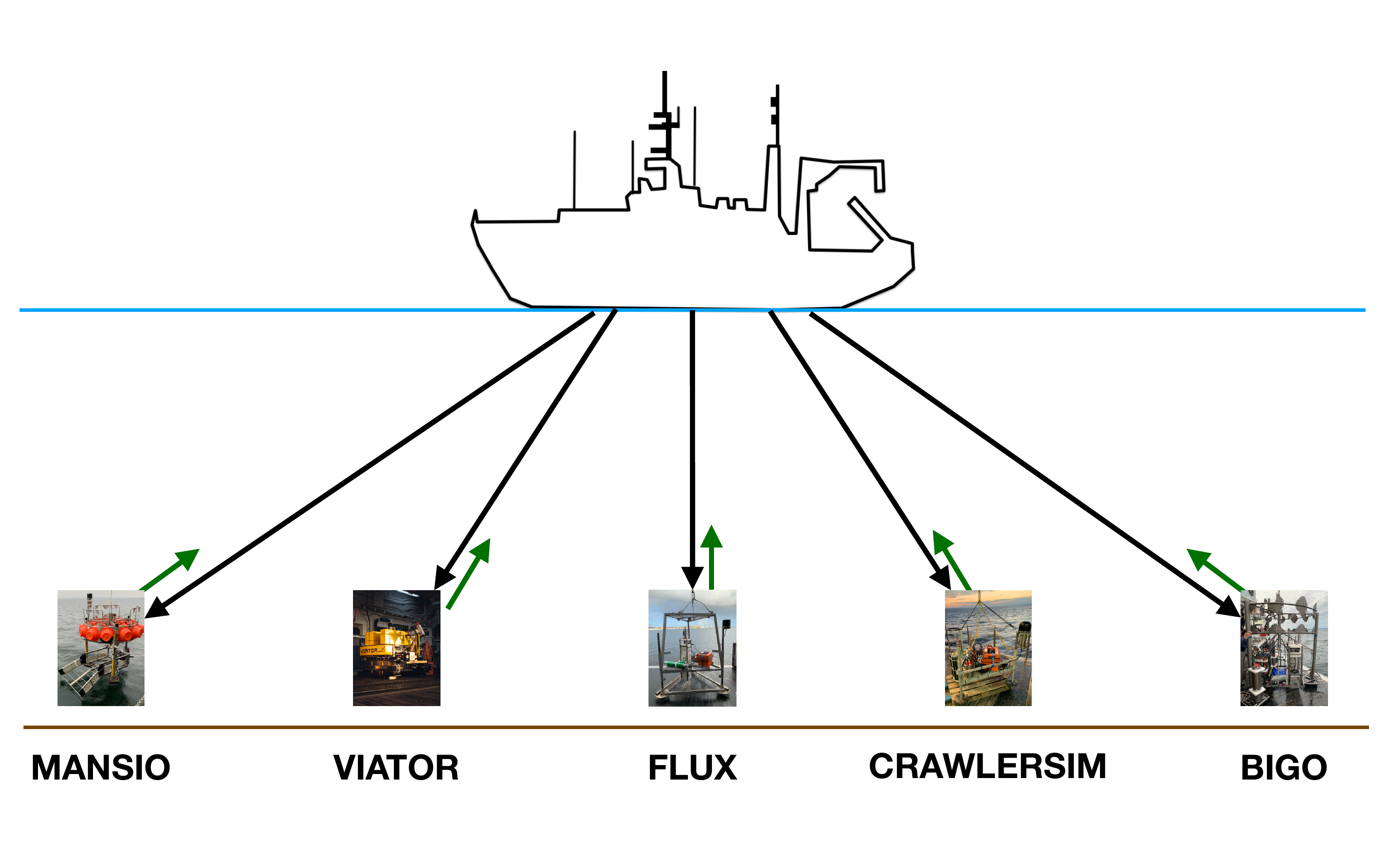}
 \caption{Synchronize behaviors with Physical Twins.}
 \label{fig:scenariob}
\end{subfigure}
\\
\begin{subfigure}{0.45\textwidth}
\centering
 \includegraphics[width=\textwidth]{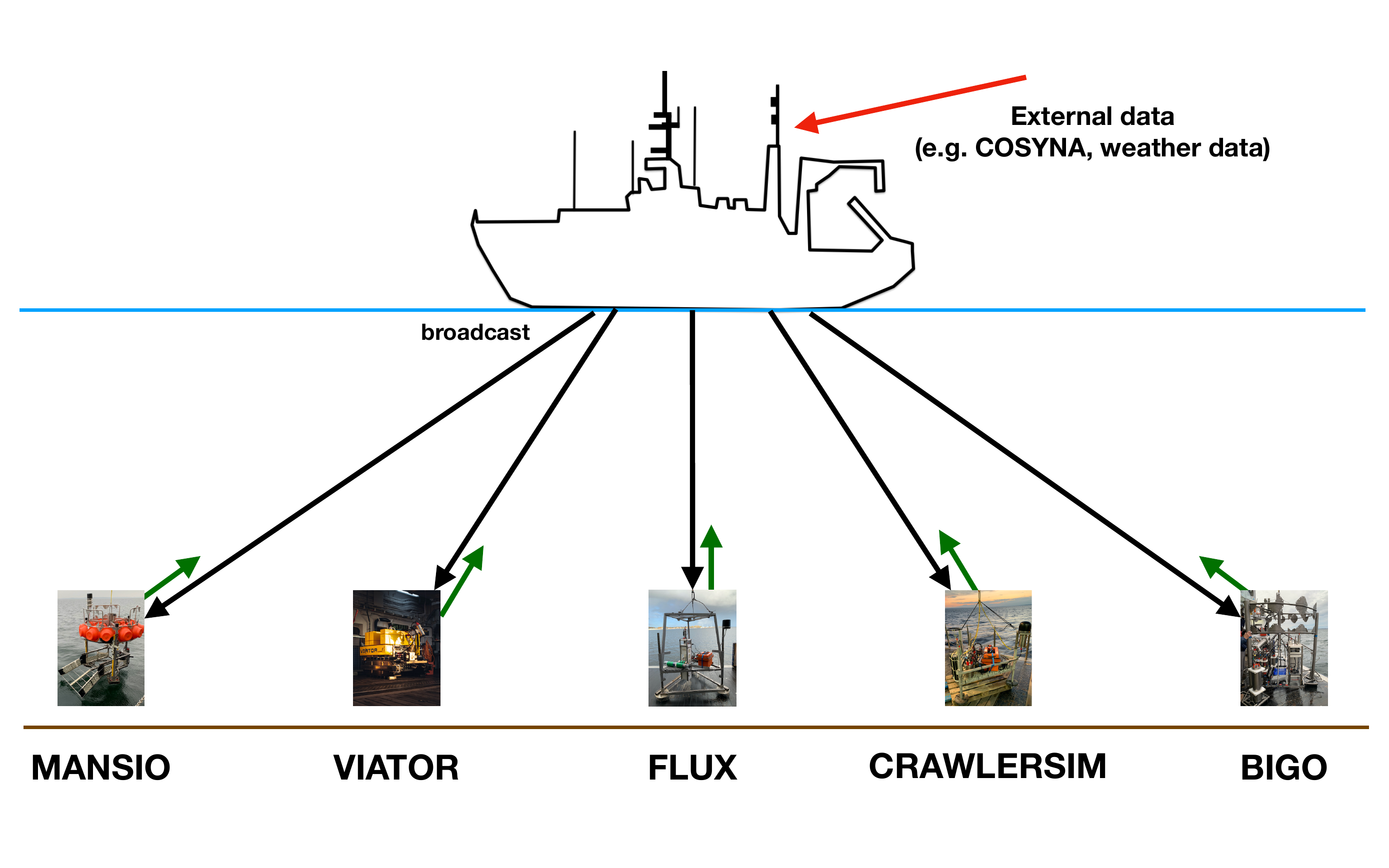}
 \caption{broadcast an event from the research vessel.}
 \label{fig:scenarioc}
\end{subfigure}
\hspace{1cm}
\begin{subfigure}{0.45\textwidth}
\centering
 \includegraphics[width=\textwidth]{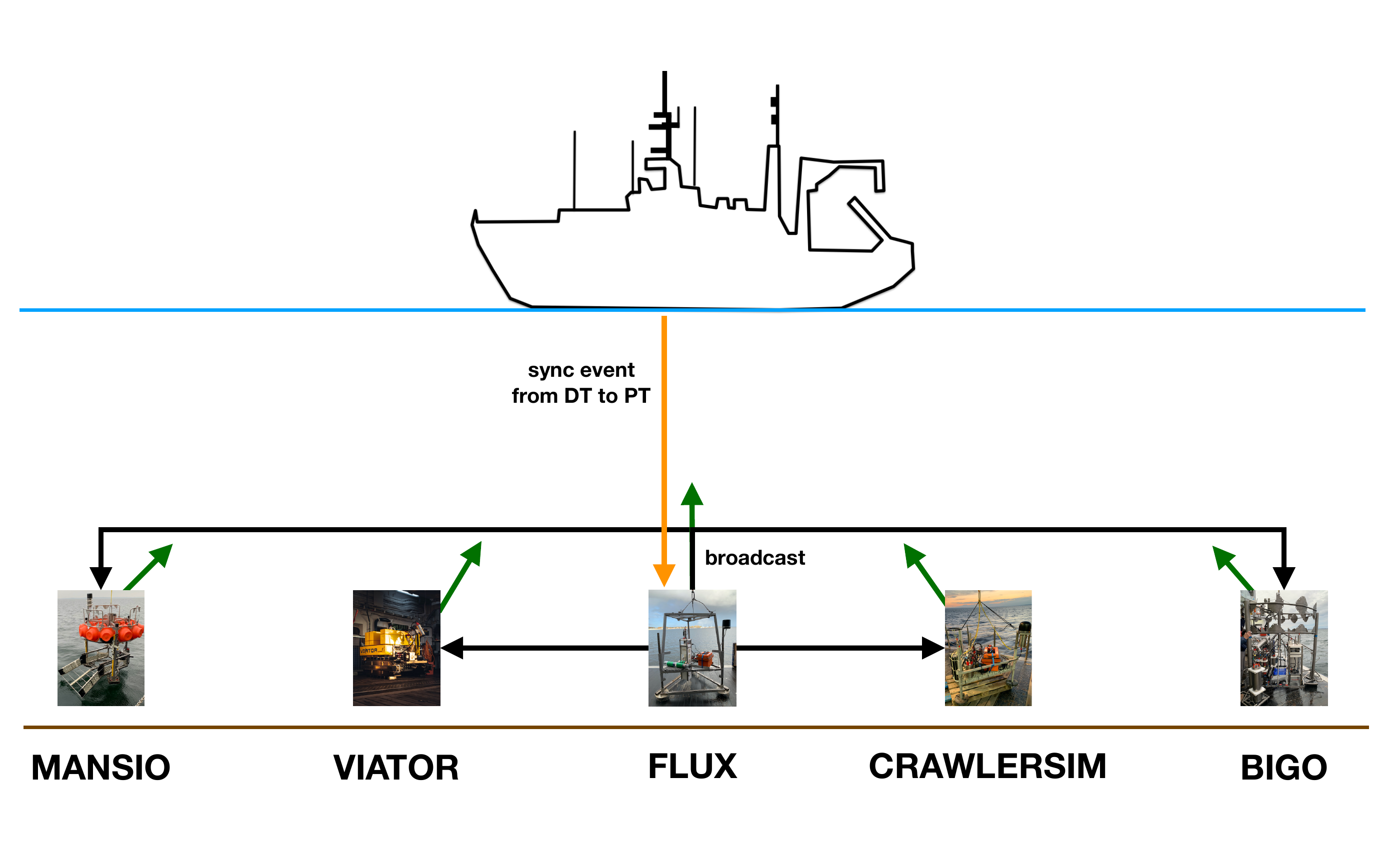}
 \caption{One Digital Twins detects an event and its Physical Twin broadcasts this event to all other platforms.}
 \label{fig:scenariod}
\end{subfigure}
\caption{Scenarios that are evaluated in the ARCHES Demo Mission. Refer to Fig.~\ref{fig:multiplelander} for larger pictures of the ocean observation system components.}
\label{fig:scenarios}
\end{figure*}
To investigate the suitability of the Digital Twin approach for under water networks, the scenarios displayed in Fig.~\ref{fig:scenarios} were evaluated. The approach was considered feasible, if all the scenarios can be repeated independently and the ocean observation systems synchronize environmental data and statuses reliably. While Scenarios \ref{fig:scenarioa} - \ref{fig:scenarioc} evaluate different aspects of the twin synchronization, Scenario \ref{fig:scenariod} evaluates (i.) the collaboration between a Physical Twin and its corresponding Digital Twin and (ii.) the communication between two Physical Twins. As a common standard between all ocean observation systems, four basic messages are exchanged: StandardO2, StandardStatus, SetBehavior, and O2Event. StandardO2 messages carry a timestamp, the oxygen value, oxygen saturation and the temperature. StandardStatus messages carry a timestamp, the current behavior ID and the status of the behavior such as running, finished, or failure. SetBehavior messages carry the ID of the behavior to execute. O2Event messages carry a string that indicates a detected environmental event. In this Demonstration Mission two event-scenarios were tested, ``Oxia'', chosen for well ventilated conditions in the ambient water body of a respective platform and ``Hypoxia'', indicating adverse environmental conditions with reduced oxygen levels.

In Scenario \ref{fig:scenarioa} all ocean observation systems were placed at the seafloor and synchronize their measurements to the corresponding Digital Twins on the research vessel. The Physical Twins are programmed to different measurement cycles from every five seconds to every five minutes, which were automatically started during the startup.

In Scenario \ref{fig:scenariob} commands were manually sent to the Digital Twin on the research vessel and from there synchronized to the corresponding Physical Twin. The Physical Twin respond with a status message of a changed behavior. MANSIO, VIATOR, and BIGO also respond with a status message periodically or after a given behavior has finished.

In Scenario \ref{fig:scenarioc} oxygen data measured by the platforms were used to identify changes in ambient oxygen levels, which were categorized either as oxic (fully ventilated) or hypoxic conditions (oxygen level declines below a certain threshold). Via broadcast an oxygen event such as Oxia or Hypoxia, was send to the other platforms, which changed their measurement strategy according to a predefined protocol. In this experiment the broadcast was manually triggered on the Basestation, without any automated decision algorithm.

Scenario \ref{fig:scenariod} combines vertical and horizontal communication from Physical Twin to Physical Twin. Similar to Scenario Fig.~\ref{fig:scenarioc} oxygen measurements were used to recognize oxic or hypoxic conditions and to trigger the corresponding platform specific measurement strategies Oxia or Hypoxia. However, the events were not broadcast from the research vessel, instead, the decision was triggered one of the Digital Twins. The Digital Twin then synchronizes its behavior to its corresponding Physical Twin and this twin broadcasts this event to all other ocean observation systems. Each observation system that receives this event, switches into the Oxia or Hypoxia behavior and synchronizes its status to the corresponding Digital Twin on the research vessel. 

\subsection{RESULTS AND DISCUSSION}
It’s beyond the scope of this field report to present the detailed behavior of the different platforms during the experiments. Instead, we will focus on the main aspects with regard to the application of Digital Twins in underwater networks.

For each ocean observation system Scenario \ref{fig:scenarioa} was evaluated in the moment it was placed on the seafloor. Since each Physical Twin starts measuring on start-up, we know that the synchronization with the corresponding Digital Twin is working, if StandardO2 messages arrive in corresponding Digital Twin. 
\begin{figure}[tb]
 \centering
 \includegraphics[width=0.5\textwidth]{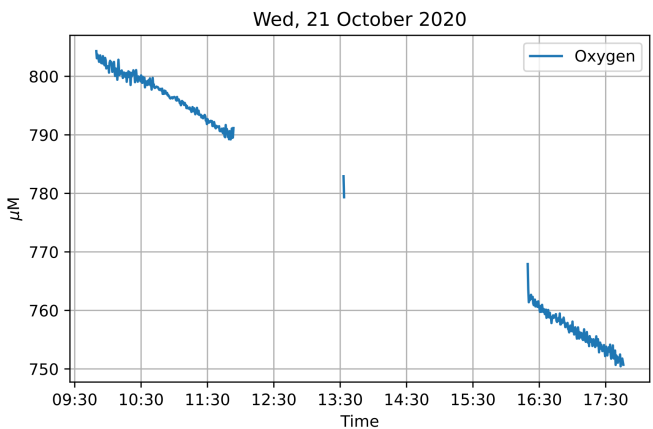}
 \caption{Incorrect oxygen measurements taken by MANSIO.}
 \label{fig:mansioo2}
\end{figure}
Scenario \ref{fig:scenariob} was executed one by one for each ocean observation system. To change the current behavior, we sent SetBehavior messages with different behaviorIDs. This command can only be send to the Digital Twins, which then synchronizes the message to the same topic on the corresponding Physical Twin. The Physical Twin solely reacts to SetBehavior messages sent by its Digital Twin. In this mission all SetBehavior commands sent to a Digital Twin were automatically synchronized to its corresponding Physical Twin. Nevertheless, this architecture allows to add an extra layer that checks, if the behavior was changed on the Digital Twin successfully and only after a successful behavior change the SetBehavior message will be synchronized to the Physical Twin. A behavior change on the Physical Twin leads to a StandardStatus message that is synchronized to its corresponding Digital Twin. In this scenario, we received this StandardStatus messages from all ocean observation systems.

Fig.~\ref{fig:mansioo2} shows a part of the results for Scenario \ref{fig:scenarioa} and \ref{fig:scenariob} for the ocean observation system MANSIO. In Fig.~\ref{fig:mansioo2} the measurements taken by MANSIO over a period of $9$ hours during the first deployment are visualized. The two gaps indicate where the Digital Twin successfully stopped and started the measurement on its physical counterpart. Notice, this measurements ($y\geq750 \, \si{\mu M}$) are contrary to measurements from the actual oxygen value during that period. The other oxygen sensors measured an oxygen concentration in the air around $\SI{230}{\mu M}$ and much less underwater. When MANSIO was deployed the first time, we forgot to remove the protective cap of its oxygen sensor, which caused this incorrect measurements. In a traditional approach, where the ocean observation system is deployed on the seafloor for a couple of weeks, this error would have been noticed only after the mission. 

In Scenario \ref{fig:scenarioc} an O2Event message was broadcast to all ocean observation systems. Upon successful reception of the message, all of them switched to their platform specific behavior, i.e. to the behavior, which has been defined for the environmental condition Oxia or Hypoxia. For example, MANSIO turned on its lights and VIATOR moved backwards\footnote{We uploaded a video with this submission}. Nevertheless, the message was not received by all ocean observation systems during all broadcasting attempts. Problems of data transmission were caused by increased distances between the RV ALKOR or unfavorable positions of the research vessel within the network area. In some cases turbulence in the water induced by the propeller of the ship might have further contributed to disturb data transmission. 

Instead of broadcasting a message from the research vessel to all the ocean observation system, an O2Event message was published to a decision node on one of the Digital Twins to evaluate the Scenario \ref{fig:scenariod}. In this scenario, the O2Event was sent to the Digital Twin of the FLUX Lander and its Physical Twin broadcast the event to all other platforms. The behaviors to be executed and results were the same as in Scenario \autoref{fig:scenarioc}. Again, not all ocean observation systems received always all broadcast messages. In horizontal communication the positioning of the ocean observation systems to each other is important. Nevertheless, this scenario showed a cooperation between a Digital Twin and its physical counterpart (Stage IV Digital Twin \cite{dtdefinition}). Since the battery capacity on ocean observation system is limited, see \autoref{sec:context}, tasks with extensive power consumption could be sourced out to the Digital Twin that is running on a server, e.g. placed on the research vessel. Power consuming tasks that are envisioned on the ocean observation systems in the context of the project ARCHES are machine learning algorithms to predict upcoming environmental changes. Velasco-Montero et al. analyzed different frameworks used for machine learning algorithms and showed the increased power consumption on a RaspberryPi Model B \cite{kipowercon}.

\section{CONCLUSION AND FUTURE WORK}\label{sec:conclusion}
In conclusion, the presented scenarios were evaluated successfully. Hence, Digital Twins Prototypes are capable to be used in underwater networks. Especially in 2020, during the travel restrictions caused by the COVID19 pandemic, the advantage of this approach was the independence of a permanent physical connection to the ocean observation systems from AWI and GEOMAR to develop and test the network. It would not have been possible to evaluate the network without this approach.

Monitoring and operating an ocean observation system from a research vessel is possible without Digital Twins. The advantage of Digital Twins is the visual response we get when synchronizing commands from the Digital Twin to the Physical Twin. If a command causes an error on the Digital Twin, it also causes an error on the Physical Twin. Hence, if an error is thrown, we immediately know where and maybe why it occurred. Adding a layer that prevents the synchronization of commands that cause errors on the Digital Twin, reduces the impact of mistakes done by an operator of that ocean observation system. Combined with a simulation tool like Gazebo, see also \autoref{sec-dtp}, different commands and scenarios can be simulated, before executed on the ocean observation system.

Furthermore, the importance of extensive software and hardware testing in the embedded domain became apparent. With the increasing complexity of the software, software tests become a prerequisite in the development process of ocean observation systems. The Digital Twin Prototype approach developed in this project allows state of the art software testing in the embedded software domain, including continuous integration~\cite{shahin_continuous_2017}. This is a generic approach, suitable for all kinds of Digital Twins, not only in the underwater domain.

The presented results for MANSIO in Scenario \ref{fig:scenarioa} hint another use case for Digital Twins: predictive maintenance. By monitoring the performance and condition of a Physical Twin's hardware and equipped with algorithms to analyze anomalies or deviations from monitored data in previous missions, a Digital Twin helps to detect failing hardware. The ocean observation system can be maintained and the failing hardware replaced before the gathered data during the entire research cruise is corrupted. Kapteyn et al. demonstrate a Digital Twin for predictive maintenance for unmanned aerial vehicles \cite{dtwillcox}.

Digital Twins do not have to run on a server on a research vessel near the ocean observation systems. Instead of expensive research cruises to retrieve the collected data from the ocean observation systems, a buoy at the ocean surface can be the gateway station between the ocean observation systems at the seafloor and a satellite. Scientists and technicians would be able to retrieve the data and operate ocean observation systems from any place in the world via the Internet. 

Continuous long-term and interconnected ocean observation will be become increasingly important when facing the climate changes. A possible context for our approach could be the already existing FRAM Ocean Observing System \cite{fram} infrastructure operated by AWI, which targets the gateway between the North Atlantic and the Central Arctic.
 
\section*{Acknowledgment}
The project is supported through the HGF-Alliance ARCHES -- Autonomous Robotic Networks to Help Modern Societies and the Helmholtz Association.

\bibliographystyle{IEEEtran}
\bibliography{IEEEabrv, ICSI-2020-11-0133.R1_Barbie-bibliography}

\begin{thebibliography}{10}
\providecommand{\url}[1]{#1}
\csname url@samestyle\endcsname
\providecommand{\newblock}{\relax}
\providecommand{\bibinfo}[2]{#2}
\providecommand{\BIBentrySTDinterwordspacing}{\spaceskip=0pt\relax}
\providecommand{\BIBentryALTinterwordstretchfactor}{4}
\providecommand{\BIBentryALTinterwordspacing}{\spaceskip=\fontdimen2\font plus
\BIBentryALTinterwordstretchfactor\fontdimen3\font minus
  \fontdimen4\font\relax}
\providecommand{\BIBforeignlanguage}[2]{{%
\expandafter\ifx\csname l@#1\endcsname\relax
\typeout{** WARNING: IEEEtran.bst: No hyphenation pattern has been}%
\typeout{** loaded for the language `#1'. Using the pattern for}%
\typeout{** the default language instead.}%
\else
\language=\csname l@#1\endcsname
\fi
#2}}
\providecommand{\BIBdecl}{\relax}
\BIBdecl

\bibitem{manets_survey_2018}
B.~U. Islam, R.~F. Olanrewaju, F.~Anwar, A.~R. Najeeb, and M.~Yaacob, ``A
  {Survey} on {MANETs}: {Architecture}, {Evolution}, {Applications}, {Security}
  {Issues} and {Solutions},'' \emph{Indonesian Journal of Electrical
  Engineering and Computer Science}, vol.~12, no.~2, pp. 832--842, 2018.

\bibitem{breitburg2018declining}
D.~Breitburg, L.~A. Levin, A.~Oschlies, M.~Gr{\'e}goire, F.~P. Chavez, D.~J.
  Conley, V.~Gar{\c{c}}on, D.~Gilbert, D.~Guti{\'e}rrez, K.~Isensee
  \emph{et~al.}, ``{Declining oxygen in the global ocean and coastal waters},''
  \emph{Science}, vol. 359, no. 6371, 2018.

\bibitem{MFI2020}
A.~Barbie, W.~Hasselbring, N.~Pech, S.~Sommer, S.~Fl{\"o}gel, and
  F.~Wenzh{\"o}fer, ``{Prototyping Autonomous Robotic Networks on Different
  Layers of RAMI 4.0 with Digital Twins},'' in \emph{Proceedings of the 2020
  IEEE International Conference on Multisensor Fusion and Integration for
  Intelligent Systems (MFI 2020)}.\hskip 1em plus 0.5em minus 0.4em\relax IEEE,
  2020, pp. 1--6.

\bibitem{sommer2009seabed}
S.~Sommer, P.~Linke, O.~Pfannkuche, T.~Schleicher, J.~S.~v. Deimling, A.~Reitz,
  M.~Haeckel, S.~Fl{\"o}gel, and C.~Hensen, ``{Seabed methane emissions and the
  habitat of frenulate tubeworms on the Captain Arutyunov mud volcano (Gulf of
  Cadiz)},'' \emph{Marine Ecology Progress Series}, vol. 382, pp. 69--86, 2009.

\bibitem{bigo2003}
O.~Pfannkuche and P.~Linke, ``{GEOMAR Landers as Long-Term Deep-Sea
  Observatories},'' \emph{Sea Technology}, vol.~44, no.~9, pp. 50--55, 2003.

\bibitem{wenzhofer2016robex}
F.~Wenzh{\"o}fer, T.~Wulff, S.~Floegel, S.~Sommer, and C.~Waldmann,
  ``{ROBEX-Innovative Robotic Technologies for Ocean Observations, a Deep-Sea
  Demonstration Mission},'' in \emph{OCEANS 2016 MTS/IEEE Monterey}.\hskip 1em
  plus 0.5em minus 0.4em\relax IEEE, 2016, pp. 1--8.

\bibitem{lemburg2018benthic}
J.~Lemburg, F.~Wenzh{\"o}fer, M.~Hofbauer, P.~F{\"a}rber, and V.~Meyer,
  ``{Benthic Crawler NOMAD},'' in \emph{2018 OCEANS-MTS/IEEE Kobe Techno-Oceans
  (OTO)}.\hskip 1em plus 0.5em minus 0.4em\relax IEEE, 2018, pp. 1--7.

\bibitem{evologicsmodem717}
\BIBentryALTinterwordspacing
{{Evologics GmbH}}, ``{{S2C R 7/17 USBL Communication and Positioning
  Device}}.'' [Online]. Available:
  \url{https://evologics.de/acoustic-modem/7-17/usbl-serie}
\BIBentrySTDinterwordspacing

\bibitem{underwaterchallenge}
I.~F. Akyildiz, D.~Pompili, and T.~Melodia, ``{{Underwater Acoustic Sensor
  Networks: Research Challenges}},'' \emph{Ad hoc networks}, vol.~3, no.~3, pp.
  257--279, 2005.

\bibitem{dtdefinition}
R.~Saracco, ``{Digital Twins: Bridging Physical Space and Cyberspace},''
  \emph{Computer}, vol.~52, no.~12, pp. 58--64, 2019.

\bibitem{rospaper}
M.~Quigley, K.~Conley, B.~Gerkey, J.~Faust, T.~Foote, J.~Leibs, R.~Wheeler, and
  A.~Y. Ng, ``{ROS: An open-source Robot Operating System},'' in \emph{ICRA
  workshop on open source software}, vol.~3, no. 3.2.\hskip 1em plus 0.5em
  minus 0.4em\relax Kobe, Japan, 2009, p.~5.

\bibitem{gazebopaper}
N.~Koenig and A.~Howard, ``{Design and Use Paradigms for Gazebo, an Open-Source
  Multi-Robot Simulator},'' in \emph{2014 IEEE/RSJ International Conference on
  Intelligent Robots and Systems (IROS) (IEEE Cat. No. 04CH37566)},
  vol.~3.\hskip 1em plus 0.5em minus 0.4em\relax IEEE, 2004, pp. 2149--2154.

\bibitem{GITEWS}
E.~R. Flueh, T.~Schoene, and W.~Weinrebe, ``{FS Sonne Fahrtbericht/Cruise
  Report SO 186 B, C \& D},'' 2005,
  \url{http://hdl.handle.net/11858/00-1735-0000-0001-3307-F}.

\bibitem{dart}
F.~Gonzalez, H.~Milburn, E.~Bernard, and J.~Newman, ``{Deep-ocean assessment
  and reporting of tsunamis (DART): Brief overview and status report},'' in
  \emph{Proceedings of the International Workshop on Tsunami Disaster
  Mitigation}, vol.~19.\hskip 1em plus 0.5em minus 0.4em\relax NOAA Tokyo,
  Japan, 1998.

\bibitem{dccl}
T.~Schneider and H.~Schmidt, ``{The Dynamic Compact Control Language: A Compact
  Marshalling Scheme for Acoustic Communications},'' in \emph{OCEANS'10 IEEE
  SYDNEY}.\hskip 1em plus 0.5em minus 0.4em\relax IEEE, 2010, pp. 1--10.

\bibitem{nasadt2012}
E.~Glaessgen and D.~Stargel, ``{The digital twin paradigm for future NASA and
  US Air Force vehicles},'' in \emph{53rd AIAA/ASME/ASCE/AHS/ASC Structures,
  Structural Dynamics and Materials Conference}, 2012, p. 1818.

\bibitem{kipowercon}
D.~{Velasco-Montero}, J.~{Femández-Bemi}, R.~{Carmona-Gálán}, and
  A.~{Rodríguez-Vázquez}, ``{On the Correlation of CNN Performance and
  Hardware Metrics for Visual Inference on a Low-Cost CPU-based Platform},'' in
  \emph{2019 International Conference on Systems, Signals and Image Processing
  (IWSSIP)}, 2019, pp. 249--254.

\bibitem{shahin_continuous_2017}
M.~Shahin, M.~Ali~Babar, and L.~Zhu, ``Continuous {Integration}, {Delivery} and
  {Deployment}: {A} {Systematic} {Review} on {Approaches}, {Tools},
  {Challenges} and {Practices},'' \emph{IEEE Access}, vol.~5, pp. 3909--3943,
  2017.

\bibitem{dtwillcox}
D.~Allaire, D.~Kordonowy, M.~Lecerf, L.~Mainini, and K.~Willcox,
  ``{Multifidelity DDDAS methods with application to a self-aware aerospace
  vehicle},'' \emph{Procedia Computer Science}, vol.~29, pp. 1182--1192, 2014.

\bibitem{fram}
T.~{Soltwedel}, U.~{Schauer}, O.~{Boebel}, E.~{Nöthig}, A.~{Bracher},
  K.~{Metfies}, I.~{Schewe}, A.~{Boetius}, and M.~{Klages}, ``{FRAM - FRontiers
  in Arctic Marine Monitoring Visions for Permanent Observations in a Gateway
  to the Arctic Ocean},'' in \emph{2013 MTS/IEEE OCEANS - Bergen}, 2013, pp.
  1--7.

\end{thebibliography}

\newpage
\noindent
\textbf{Alexander Barbie} is a software engineer at the Alfred-Wegener-Institute Helmholtz Centre for Polar and Marine Research (Bremberhaven, Germany) and the GEOMAR Helmholtz Centre for Ocean Research Kiel (Germany), and a doctoral researcher in the software engineering group in the Department of Computer Science, Kiel University, Germany. Contact him at abarbie@geomar.de.

\vspace{1em}
\noindent
\textbf{Niklas Pech} is a software engineer at the GEOMAR Helmholtz Centre for Ocean Research Kiel (Germany). Contact him at npech@geomar.de.

\vspace{1em}
\noindent
\textbf{Wilhelm Hasselbring} is a full professor of software engineering in the Department of Computer Science at Kiel University, Germany. Contact him at hasselbring@email.uni-kiel.de.

\vspace{1em}
\noindent
\textbf{Sascha Flögel} is a senior scientist at the GEOMAR Helmholtz Centre for Ocean Research Kiel (Germany). Contact him at sfloegel@geomar.de.

\vspace{1em}
\noindent
\textbf{Frank Wenzhöfer} is a senior scientist in the HGF-MPG Joint Research Group on Deep-Sea Ecology and Technology at the Alfred-Wegener-Institute for Polar and Marine Research, Bremerhaven, Germany and is associated scientist at the Max Planck Institute for Marine Microbiology, Bremen, Germany. Contact him at f.wenzhoefer@awi.de.

\vspace{1em}
\noindent
\textbf{Michael Walter} is a postdoctoral scientist in the Institute of Communications and Navigation at the German Aerospace Center, Oberpfaffenhofen-Wessling, Germany. Contact him at m.walter@dlr.

\vspace{1em}
\noindent
\textbf{Elena Shchekinova} is a postdoctoral scientist at the GEOMAR Helmholtz Centre for Ocean Research Kiel (Germany). Contact her at eshchekinova@geomar.de.

\vspace{1em}
\noindent
\textbf{Marc Busse} is an electrical engineer at the GEOMAR Helmholtz Centre for Ocean Research Kiel (Germany). Contact him at mbusse@geomar.de.

\vspace{1em}
\noindent
\textbf{Matthias T\"urk} is an electrical engineer at the GEOMAR Helmholtz Centre for Ocean Research Kiel (Germany). Contact him at mtuerk@geomar.de.

\vspace{1em}
\noindent
\textbf{Michael Hofbauer} is an electrical engineer in the HGF-MPG Joint Research Group on Deep-Sea Ecology and Technology at the Alfred-Wegener-Institute for Polar and Marine Research, Bremerhaven, Germany and at the Max Planck Institute for Marine Microbiology, Bremen, Germany. Contact him at m.hofbauer@awi.de.

\vspace{1em}
\noindent
\textbf{Stefan Sommer} is a senior scientist at the GEOMAR Helmholtz Centre for Ocean Research Kiel (Germany). Contact him at ssommer@geomar.de.
\end{document}